%% file: main.tex
\documentclass[USenglish,twocolumn]{article}
\usepackage[utf8]{inputenc}
\usepackage[big,online]{dgruyter}
\usepackage[sort&compress, square, numbers]{natbib}
\usepackage{times}
\usepackage{graphicx} 
\usepackage{amsmath}
\usepackage{psfrag}
\usepackage{siunitx}
\usepackage{color}
\usepackage{tikz}
\usepackage{pgfplots}
\pgfplotsset{compat=1.18}
\usepackage[dvipsnames,table]{xcolor}
\usepackage{float} 
\usepackage{tocloft}
\usepackage{wrapfig}
\usepackage{subcaption}
\usepackage{comment}

\usepackage{algorithmic}
\usepackage[normalem]{ulem}

\def\showcorrections{false}

\ifthenelse{\equal{\showcorrections}{true}}{
    
}{
    
}

\begin{document}
  \DOI{10.1515/}
  \openaccess
  \pagenumbering{gobble}

\title{Towards a Pseudo-Labeling Workflow for Celltype-Classification from Explanted Brain Slice Recordings}
\runningtitle{Towards a Pseudo-Labeling Workflow for Celltype-Classification from Explanted Brain Slice Recordings}

\author*[1]{Cora~Jostock}
\author[3]{Jonas~Ort}
\author[2]{Henner~Koch}
\author[1]{Gregor~Schiele} 
\author[1]{Andreas~Erbslöh} 
\runningauthor{C.~Jostock et al.}

\affil[1]{\protect\raggedright
  University of Duisburg-Essen, Intelligent Embedded Systems Lab, Duisburg, Germany}
\affil[2]{\protect\raggedright 
  RWTH University Hospital, Department of Epileptology, Aachen, Germany}
\affil[3]{\protect\raggedright 
  RWTH University Hospital, Department of Neurosurgery, Aachen, Germany}
	
\abstract{This paper proposes an unsupervised workflow to pseudo-label extracellular spikes from human brain slice MEA recordings into two putative cell types: pyramidal cells and interneurons. Here, the raw data from the data acquisition system is used and processed. The pipeline for pre-processing includes bandpass filtering, threshold-based spike detection, frame alignment and normalization. In the ML workflow, dimensionality reduction (PCA, t-SNE, UMAP), clustering (GMM, k-means). To achieve an online system, template matching and OSort under varying curation strictness is also considered. All pipelines are evaluated by different cluster quality with within-cluster Pearson correlation, Silhouette score, and Calinski–Harabasz index. Applying stricter curation improves separation at some cost to inclusivity.
}
\keywords{brain slices, extracellular recordings, end-to-end processing, machine learning, spike sorting}

\maketitle
\input{body/0_introduction}
\input{body/1_methods}
\input{body/2_results}
\input{body/3_ending}
\vspace*{-3mm}
\bibliographystyle{IEEEtran}
\bibliography{references}

\end{document}

%% file: body/0_introduction.tex
\vspace*{-2mm}
\section{Introduction}
\vspace*{-2mm}
Neurodegenerative diseases can lead to sensory and motor impairments in humans. Invasive Brain–computer interfaces (BCIs) can partially restore these functions using microelectrode arrays~(MEAs) as electrode–tissue interface. Thus, electrical stimulation is used to impart information whilst simultaneously measuring neural activity~\cite{Erbsloeh2024, Lee2021}. Understanding the disease mechanisms in the affected tissue and translating this knowledge into a technical system may allow adaptive stimulation to provide stable long-term therapeutic efficacy~\cite{Belkacem2023ClosedLoopBCI}. 

Epilepsy poses a major societal challenge due to the unpredictability of seizures, limiting patients’ autonomy, safety, and social participation while placing a significant burden on healthcare systems~\cite{WHO2019Epilepsy}. Seizures are commonly attributed to an imbalance between excitation and inhibition within neural networks, leading to pathological hypersynchronization and recurrent disruptions of normal brain activity~\cite{Yizhar2011EIBalance}. At cellular level, this process is primarily governed by excitatory pyramidal cells and inhibitory interneurons. Distinguishing their activity from extracellular spike recordings is therefore essential for understanding seizure mechanisms and enabling targeted therapeutic interventions~\cite{Bartho2004Characterization}.
\begin{figure}[t]   \centering
    \vspace*{-1mm}
    \includegraphics[width=0.99\linewidth, clip]{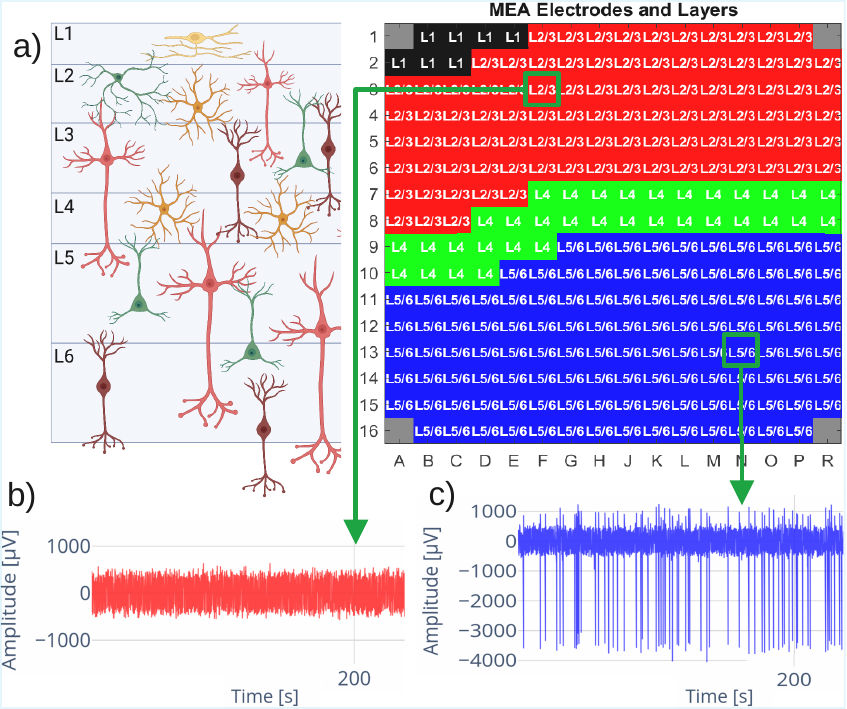}
    \caption{\textbf{a)} left: Composition of human neocortex with pyramidal cells (red), interneurons (green). right: Layout of the MCS 256MEA electrode array with definition of the brain layer of a recording  (black: Layer 1, red: Layer 2/3, green: Layer 4, blue: Layer 5/6, grey: no electrode). \textbf{b)} Neural activity on channel G3 (layer 2/3). \textbf{c)} Neural activity on channel N13 (Layer 5/6).}
    \label{fig:signals_across_layers}
    \vspace*{-2mm}
\end{figure}

\begin{figure*}[h]    \centering
    \includegraphics[width=0.92\linewidth, clip]{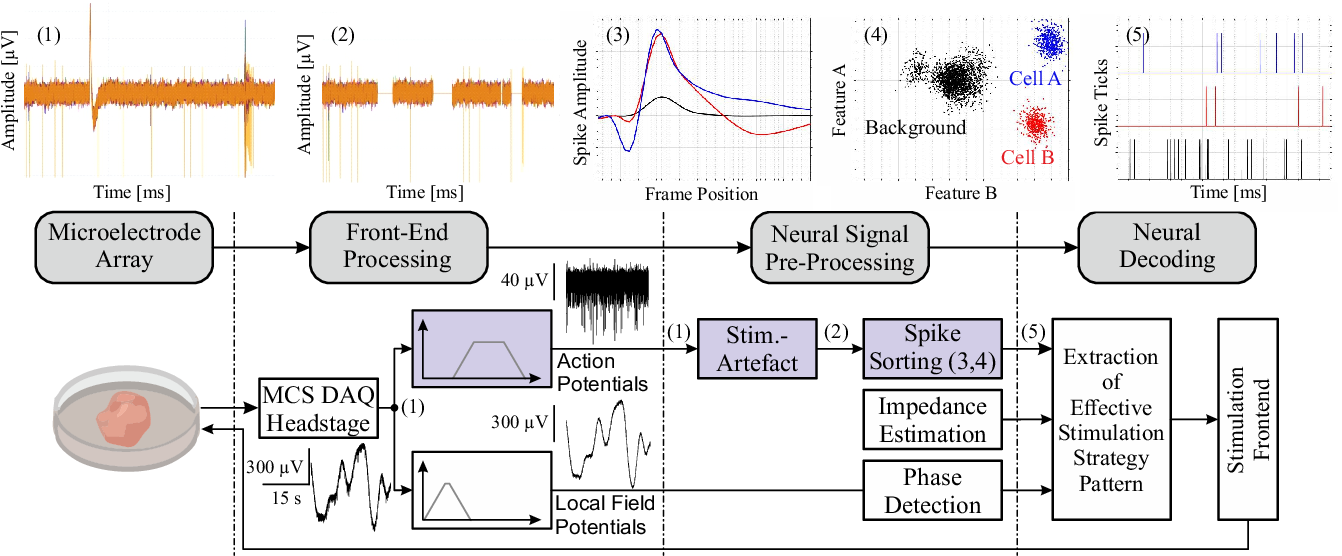}
    \caption{Concept of an end-to-end neural signal processing pipeline for explanted brain slice experiments: i) explanting the brain slice and placing on MCS 256MEA200/30iR-ITO, ii) recording the neural activity using the MCS USB-MEA256 data acquisition~(DAQ) headstage, iii) digital processing with bandpass-filters and stimulation artefact remover for extracting clean spike activity, iv) applying ML model to extract specific cell-types and iv) decode the neural behavior / strategy to perform closed-loop stimulation. Corresponding examples of the signals are given and the purple highlighted boxes indicates the described workflow within this paper.}
    \label{fig:pipeline}
    \vspace*{-3mm}
\end{figure*}

This paper presents a machine learning (ML)-based workflow to derive pseudo-labels for pyramidal cell and interneuron classification using spike waveforms in extracellular MEA recordings on explanted brain slices. We determine the quality performance of different approaches and assess online feasibility through data curation.
A key challenge in brain-slice MEA recordings is that stimulation- and acquisition-related artifacts (e.g. activity from the other cell types) can distort the raw time series and thereby affect spike detection, waveform shapes, and stability of clustering. 
Consequently, pseudo-labels inferred from the recordings may be highly sensitive to the degree of artifact removal and event-level quality curation. Since the impact of artifact curation is difficult to quantify, we assess how different levels of “curation strictness” influence clustering metrics and the online feasibility~\cite{Erbsloeh2024}.

%% file: body/1_methods.tex
\vspace*{-2mm}
\section{Methods}
\vspace*{-2mm}
This section presents i) the structure of the extracellular recordings with the time series raw data, ii) the pre-processing methods to build a data~set and iii) the ML workflow and the used metrics.
\vspace*{-5mm}
\subsection{Data Acquisition (DAQ)} 
\vspace*{-2mm}
The data of the brain slice recordings is provided by the Koch Group (Department of Neurology, Section Epileptology at the University Hospital RWTH Aachen, Germany) and the tissue is prepared using the protocol in~\cite{Bak2024}. The data is acquired with the USB-MEA256 system using the 256MEA200/30iR-ITO array (16$\times$16 grid, \SI{30}{\micro m} electrode diameter, \SI{200}{\micro m} electrode spacing) from Multichannel Systems GmbH (MCS) at a sampling rate of~\SI{10}{kHz}, an input voltage range of~\SI{\pm3.7}{mV} @ 16-bit resolution (minimal voltage change of~\SI{112,92}{nV}) and an input bandpass filter of~\SI{1}{Hz}-\SI{5}{kHz}. Artificial cerebrospinal fluid (aCSF) was used as the recording medium to preserve physiological extracellular conditions during measurements.
\vspace*{-5mm}
\subsection{Used Data Structure and Conditions}
\vspace*{-2mm}
Figure~\ref{fig:signals_across_layers} shows the relationship between the layers from the cross-section of the brain slice (a), placed on the 256MEA with 252 active channels and four reference/control channels (b). Due to the nature of the surface electrodes, the recording quality depends on tissue condition, tissue-electrode distance, and electrode impedance~\cite{Pettersen2012extracellular,Erbsloeh2024}.
The human neocortex comprises six layers with distinct cell-type compositions and activity patterns. Pyramidal cells and interneurons are present across all layers, but L5/6 contain the highest density of large pyramidal cells and fast-spiking interneurons, exhibiting markedly stronger spiking than superficial layers~\cite{Bartho2004Characterization,MaoStaiger2024}. Therefore, analyses focus on channels covering L5/6, where a two-cluster solution is anticipated to yield~$\leq$\SI{75}{\%} for the pyramidal and~$\geq$\SI{25}{\%} for the interneuron cluster. Figure~\ref{fig:signals_across_layers} b,c) indicates sparse activity in L2/3 compared to L5/6. 
\vspace*{-4mm}
\subsection{Pre-Processing and ML-Workflow} 
\vspace*{-2mm}
Figure~\ref{fig:pipeline} shows the end-to-end signal processing pipeline to enable closed-loop stimulation in multi-modal manner for explanted brain slice experiments. We use the Open~Source Python~framework~\textit{denspp.offline} for neural data analysis~\cite{Buron2023}\footnote{Available on GitHub: \url{https://github.com/es-ude/denspp.offline}}.
The proposed pipeline receives the raw recorded data channelwise. A second order Butterworth band-pass~filter is applied to extract spiking activity (1-\SI{4.5}{kHz}) and artifacts are removed by blanking the signal globally (see example Figure~\ref{fig:pipeline} (1), (2))~\cite{Bartho2004Characterization,MaoStaiger2024}. From this raw data stream, the spikes are detected via amplitude thresholding. If a spike is available, then a spike frame with a window size of~\SI{1.6}{ms} (16~samples) is generated and the minimum is aligned to the timepoint at~\SI{400}{\micro s}. All spike waveforms are min-max normalized in order to neglect the impact of the tissue-electrode distance and increase the model generalization~\cite{Pettersen2012extracellular}.
\begin{figure}[ht]  \centering
    \includegraphics[width=0.95\linewidth, clip]{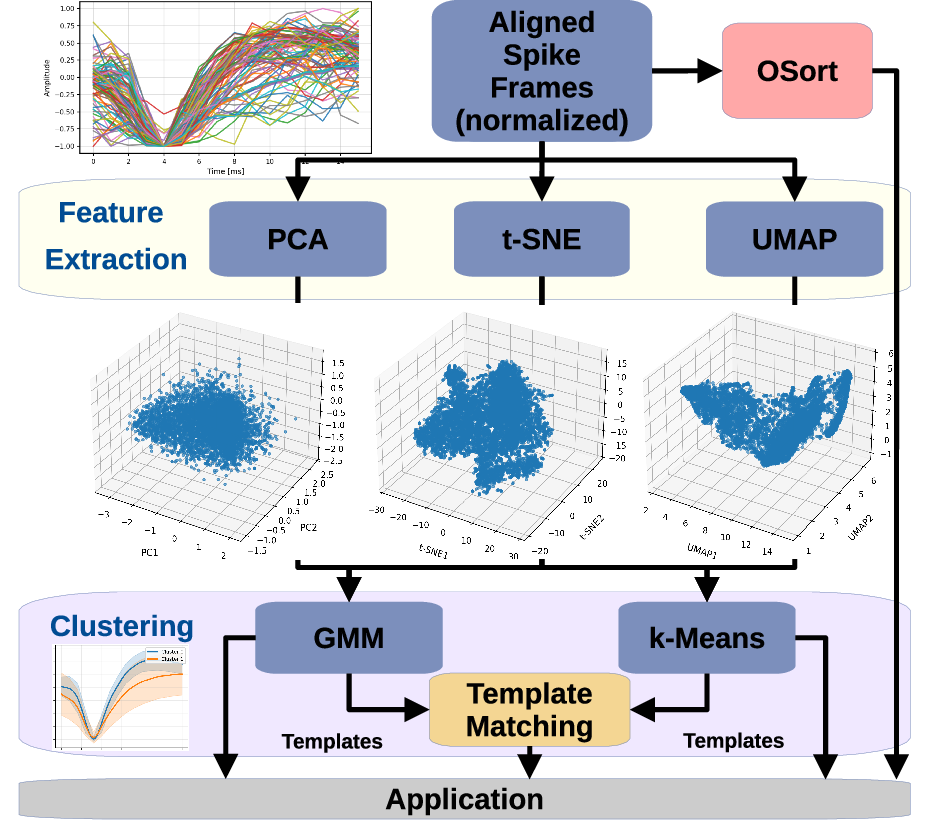}
    \caption{Clustering workflow to extract the feature space using different feature extraction and clustering methods.}
    \label{fig:ml_workflow}
    \vspace*{-4mm}
\end{figure}

This pre-processing workflow enables to build a dataset from each channel and recording in order to analyse the classification quality of the two cell types. These normalized spike frames are applied to different unsupervised ML methods in order to get the corresponding labels of the detected cell-type cluster. 
Figure~\ref{fig:ml_workflow} shows different approaches for feature extraction like Principal Component Analysis (PCA), t-Stochastic Neighbor Embedding (t-SNE) and Uniform Manifold Approximation and Projection (UMAP), in which three feature values are extracted. 
For clustering, we consider Gaussian Mixture Model (GMM) and k-Means. 

After the initial clustering, we apply an optional cleaning step of the cluster outputs (at most two iterations) if separation and proportions, also from the feature space, were suboptimal. Here, we reassigned frames to the best‑correlating centroid or removed them when both correlations fell below a fixed threshold. The labeled spike frames can go into application or to enable Template Matching with predefined clusters.

We assume that residual artifacts and low-quality events can substantially degrade clustering outcomes, especially when pseudo-labels are derived from raw recordings. To emulate artifact curation, we evaluate OSort on three frame sets (\emph{all}, \emph{majority} - frames retained by the majority of cleaned sets, \emph{strict} - frames retained by all cleaned sets), thereby quantifying how curation strictness impacts clustering quality and online robustness.

\vspace*{-4mm}
\subsection{Used Metrics}
\vspace*{-2mm}
For determining the clustering performance, we consider within‑cluster similarity by the mean pairwise Pearson correlation (zero-lag), inter- and intra-cluster distance by the Silhouette Score and Calinski-Harabasz Index (CH). The Pearson correlation is computed as the cosine between mean-centered signals within each cluster, using: 
\begin{equation}
    r(\mathbf{x},\mathbf{t})=\frac{(\mathbf{x}-\bar x\,\mathbf 1)^\top(\mathbf{t}-\bar t\,\mathbf 1)}{\|\mathbf{x}-\bar x\,\mathbf 1\|\,\|\mathbf{t}-\bar t\,\mathbf 1\|}
\end{equation}
$\mathbf{x}$: signal vector; $\mathbf{t}$: template vector; $\bar{x},\bar{t}$: sample means.

Further, we quantify cluster quality using the Silhouette score and the Calinski-Harabasz (CH) index. The Silhouette score summarizes point-wise separation vs. compactness, while CH measures the ratio of between-cluster to within-cluster dispersion. Higher values indicate better clustering in both metrics.

%% file: body/2_results.tex
\vspace*{-4mm}
\section{Results and Discussion} 
\begin{figure}[ht]    \centering
    \includegraphics[width=0.99\linewidth, clip]{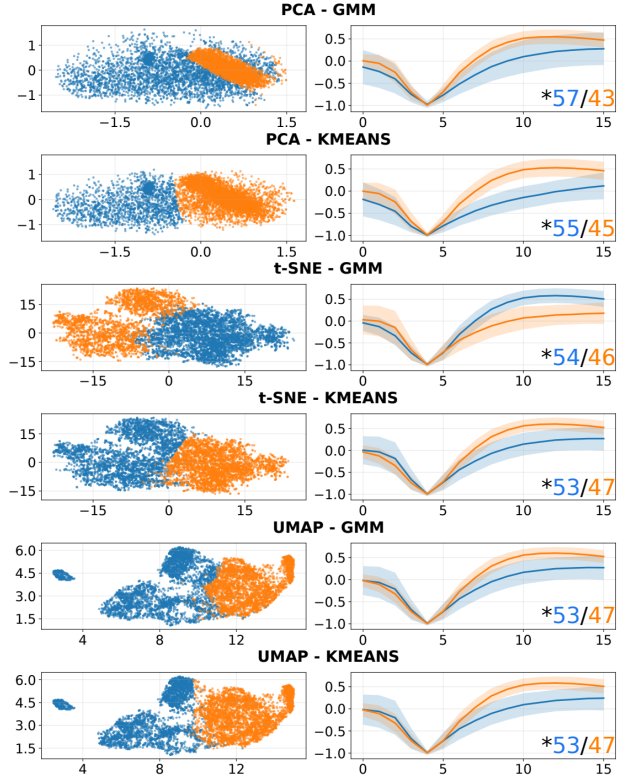}
    \caption{Representing the feature space and cluster results including *cluster proportions in percent (rounded to whole numbers) for all classical feature extraction and clustering methods after the cleaning stage. }
    \label{fig:fspace}
\end{figure}

\vspace*{-2mm}
Figure~\ref{fig:fspace} shows the feature space with building clusters (left) and the corresponding mean spike waveform of the clusters (right) for all pipelines. PCA+GMM separates less clearly than PCA+k-Means, whereas t-SNE and UMAP yield similar cluster structure for GMM and k-Means. However, cluster proportions deviate from the expected class prior for most classical pipelines.

Figure~\ref{fig:metrics} compares clustering quality across methods via metrics from section~2.4.
Among classical pipelines, t‑SNE\,+\,k‑Means achieved the highest within‑cluster similarity (0.806$\rightarrow$0.856), followed by UMAP\,+\,GMM (0.806$\rightarrow$0.854). Cleaning increased similarity consistently by $\approx$0.04–0.06 (e.g., PCA\,+\,k‑Means: 0.744$\rightarrow$0.801; t‑SNE\,+\,GMM: 0.782$\rightarrow$0.841). Template matching yields the highest within‑cluster similarity (0.85–0.87) but weak separation (CH: 1.6k–2.7k, silhouette: 0.20–0.27). OSort shows a clear similarity–separation trade‑off. \emph{Strict} attains the best separation (silhouette 0.478, CH 5.25k) with high similarity (0.838); \emph{majority} yields the highest similarity (0.869) but weak separation (silhouette 0.134, CH 1.23k); \emph{all} lies in between (similarity 0.827, silhouette 0.185, CH 2.08k). 
This pattern is consistent with our assumption that artifact-related contamination primarily harms inter-cluster separation: stricter curation (strict) removes ambiguous/contaminated events and improves separability, whereas more inclusive sets (majority/all) retain more borderline frames, boosting within-cluster similarity but weakening separation.

For application, high correlation indicates consistent waveform templates, but low silhouette/CH implies insufficient inter-cluster separation, i.e., pseudo-labels may be unstable and need further optimization (artifact/quality gating, parameters, and validation) before reliable online use. This supports that residual artifacts primarily harm separability: stricter curation removes ambiguous events and improves separation, whereas more inclusive sets retain borderline frames and blur boundaries. In addition, recordings may include spikes from other, less prevalent cell types, so enforcing exactly two clusters can merge sub-populations and reduce separability; therefore, the optimal cluster count should be estimated (e.g., elbow method and/or BIC/AIC) as part of the workflow.

\definecolor{Ocolor}{RGB}{191, 191, 191}    
\definecolor{Ccolor}{RGB}{178, 252, 178}    
\begin{figure}[t]
\begin{flushright}
\begin{tikzpicture}
\begin{axis}[
  ybar,
  bar width=6pt,                 
  width=0.85\linewidth,
  height=0.10\textwidth,
  scale only axis=true,
  trim axis left, trim axis right,
  grid=major,
  ymin=0.73, ymax=0.88,
  ytick = {0.73, 0.78, 0.83, 0.88},
  enlarge x limits={abs=0.30},
  xtick={0,1,2,3,4,5,6},
  xticklabels={},                
  xticklabel style={rotate=45, anchor=east, font=\scriptsize},
  tick label style={font=\small},
  ylabel={Correlation}
]
\addplot+[ybar, bar shift=-0.9*\pgfplotbarwidth, fill=Ocolor, draw=black!60] coordinates {
  (0,0.7801) (1,0.7442) (2,0.7818) (3,0.8062) (4,0.8057) (5,0.7938)
};
\addplot+[ybar, bar shift=0pt, fill=Ccolor, draw=black!60] coordinates {
  (0,0.8207) (1,0.8014) (2,0.8414) (3,0.8561) (4,0.8538) (5,0.8445)
};
\addplot+[ybar, bar shift=+0.9*\pgfplotbarwidth, fill=orange!40, draw=orange!70!black] coordinates {
  (0,0.8458) (1,0.8495) (2,0.8653) (3,0.8707) (4,0.8707) (5,0.8671)
};
\addplot+[ybar, bar shift=-0.9*\pgfplotbarwidth, fill=red!35, draw=red!80!black] coordinates {(6,0.8271)};
\addplot+[ybar, bar shift=0pt,              fill=red!55, draw=red!85!black] coordinates {(6,0.8686)};
\addplot+[ybar, bar shift=+0.9*\pgfplotbarwidth, fill=red!75, draw=red!90!black] coordinates {(6,0.8382)};
\end{axis}
\end{tikzpicture}
\end{flushright}
\vspace{-11mm}
\begin{flushright}
\begin{tikzpicture}
\begin{axis}[
  ybar,
  bar width=6pt,
  width=0.85\linewidth,
  height=0.10\textwidth,
  scale only axis=true,
  trim axis left, trim axis right,
  grid=major,
  ymin=500, ymax=8500,
  ytick={500,3500,6000,8500},
  yticklabels={0.5k, 3.5k,6k,8.5k},
  enlarge x limits={abs=0.30},
  xtick={0,1,2,3,4,5,6},
  xticklabels={}, 
  xticklabel style={rotate=45, anchor=east, font=\scriptsize},
  tick label style={font=\small},
  ylabel={CH-Index}
]
\addplot+[ybar, bar shift=-0.9*\pgfplotbarwidth, fill=Ocolor, draw=black!70] coordinates {
  (0,2226.1) (1,6933.3) (2,3960.3) (3,4760.1) (4,7526.4) (5,8295.1)
};
\addplot+[ybar, bar shift=0pt, fill=Ccolor, draw=green!40!black] coordinates {
  (0,2885.0) (1,7151.8) (2,3654.2) (3,3517.1) (4,5459.5) (5,6324.4)
};
\addplot+[ybar, bar shift=+0.9*\pgfplotbarwidth, fill=orange!40, draw=orange!70!black] coordinates {
  (0,1635.9828) (1,2045.4546) (2,3023.6360) (3,2826.8840) (4,2591.2417) (5,2728.4272)
};
\addplot+[ybar, bar shift=-0.9*\pgfplotbarwidth, fill=red!35, draw=red!80!black] coordinates {(6,2077.5031)};
\addplot+[ybar, bar shift=0pt,                fill=red!55, draw=red!85!black] coordinates {(6,1227.1870)};
\addplot+[ybar, bar shift=+0.9*\pgfplotbarwidth, fill=red!75, draw=red!90!black] coordinates {(6,5251.1833)};
\end{axis}
\end{tikzpicture}
\end{flushright}
\vspace{-11mm}
\begin{flushright}
\begin{tikzpicture}
\begin{axis}[
  ybar,
  bar width=6pt,
  width=0.85\linewidth,
  height=0.10\textwidth,
  scale only axis=true,
  trim axis left, trim axis right,
  grid=major,
  ymin=0.1, ymax=0.51,
  ytick = {0.1, 0.2, 0.3, 0.4, 0.5},
  enlarge x limits={abs=0.30},
  xtick={0,1,2,3,4,5,6},
  xticklabels={PCA-GMM,PCA-kMeans,t-SNE-GMM,t-SNE-kMeans,UMAP-GMM,UMAP-kMeans,OSort},
  xticklabel style={rotate=30, anchor=east, font=\scriptsize},
  tick label style={font=\small},
  ylabel={Silhouette},
  legend columns=3,
  transpose legend,         
  legend cell align=left,
  legend style={
    font=\footnotesize,
    draw=none,
    at={(0.5,-0.68)}, anchor=north, 
    column sep=12pt
  }
]
\addplot+[ybar, bar shift=-0.9*\pgfplotbarwidth, fill=Ocolor, draw=black!60] coordinates {
  (0,0.2486) (1,0.4534) (2,0.2998) (3,0.3159) (4,0.3940) (5,0.4293)
};
\addplot+[ybar, bar shift=0pt, fill=Ccolor, draw=black!60] coordinates {
  (0,0.2905) (1,0.4888) (2,0.3121) (3,0.2969) (4,0.3679) (5,0.4153)
};
\addplot+[ybar, bar shift=+0.9*\pgfplotbarwidth, fill=orange!40, draw=orange!70!black] coordinates {
  (0,0.1986) (1,0.2574) (2,0.2677) (3,0.2500) (4,0.2370) (5,0.2507)
};
\addplot+[ybar, bar shift=-0.9*\pgfplotbarwidth, fill=red!35, draw=red!80!black] coordinates {(6,0.1854)};
\addplot+[ybar, bar shift=0pt,                fill=red!55, draw=red!85!black] coordinates {(6,0.1342)};
\addplot+[ybar, bar shift=+0.9*\pgfplotbarwidth, fill=red!75, draw=red!90!black] coordinates {(6,0.4780)};
\legend{
  Original,
  Cleaned,
  Template Matching,
  OSort all,
  OSort majority,
  OSort strict
}
\end{axis}
\end{tikzpicture}
\end{flushright}
\vspace*{-4mm}
\caption{Comparison across methods (top: Correlation; middle: Calinski–Harabasz; bottom: Silhouette).}
\label{fig:metrics}
\vspace*{-4mm}
\end{figure}
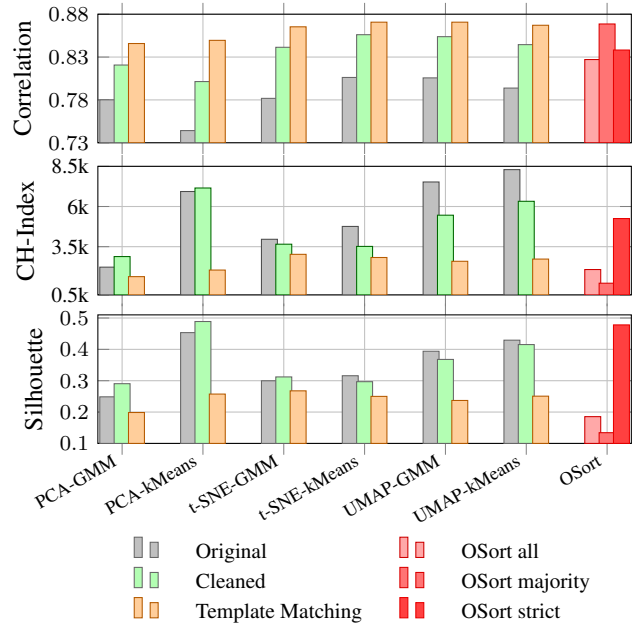

%% file: body/3_ending.tex
\vspace*{-12mm}
\section{Conclusion and Outlook} 
\vspace*{-2mm}
This paper presents a workflow in order to identify clusters from extracellular spike recordings from time-series raw data by applying unsupervised ML techniques. Here, we compared different approaches / pipelines in order to evaluate a good configuration to deploy a real‑time–capable classifier for brain slice recordings. Moreover, the evaluation shows that different feature‑extraction and clustering approaches in combination with a data curation provide complementary strengths, supporting the robustness of the overall pseudo‑labeling strategy. In addition, the right selection of the data‑curation strictness is highly sensitive on the model performance. OSort achieves a good trade-off between correlation and explainability in order to enable an online classifier.

Future work should focus on validating and optimizing the workflow across larger and more heterogeneous data~sets to increase generalizability, specially applying OSort into a real-time signal processing system. Also, establishing a reliable strategy for automatic artifact removing is necessary. With the pseudo labels, We want to train an autoencoder-based
anomaly detection model to increase explanaibility of the
OSort classifier.